\documentclass[12pt,a4paper]{article}
\usepackage{a4wide}
\usepackage{amsmath}
\usepackage{amssymb}
\usepackage{amsfonts}
\usepackage{epsfig}
\usepackage{subfigure}
\usepackage{exscale}
\usepackage{float}
\usepackage{bbm}
\usepackage[numbers,sort&compress]{natbib}

\newcommand{\CP}{{\mathbb{C}P}}

\newcommand{\p}{\partial}

\makeatletter
\@addtoreset{equation}{section}
\makeatother

\title{From an Antiferromagnet to a Valence Bond Solid: \\
Evidence for a First Order Phase Transition}

\author{F.-J.\ Jiang$^a$, M.\ Nyfeler$^a$, S.\ Chandrasekharan$^b$, and 
U.-J.\ Wiese$^a$
\\ \\
$^a$ Institute for Theoretical Physics, Bern University \\
Sidlerstrasse 5, CH-3012 Bern, Switzerland \\ \\
$^b$ Department of Physics, Box 90305, Duke University \\
Durham, North Carolina 27708 \\ \\}

\begin{document} 

\maketitle

\vspace{-1cm}

\begin{abstract} \normalsize

Using a loop-cluster algorithm we investigate the spin $\frac{1}{2}$ Heisenberg
antiferromagnet on a square lattice with exchange coupling $J$ and an 
additional four-spin interaction of strength $Q$. We confirm the existence of a
phase transition separating antiferromagnetism at $J/Q > J_c/Q$ from a valence
bond solid (VBS) state at $J/Q < J_c/Q$. Although our Monte Carlo data are
consistent with those of previous studies, we do not confirm the existence of a
deconfined quantum critical point. Instead, using a flowgram method on lattices
as large as $80^2$, we find evidence for a weak first order phase 
transition. We also present a detailed study of the antiferromagnetic phase. 
For $J/Q > J_c/Q$ the staggered magnetization, the spin stiffness, and the
spinwave velocity of the antiferromagnet are determined by fitting Monte Carlo
data to analytic results from the systematic low-energy effective field theory
for magnons. Finally, we also investigate the physics of the VBS state at 
$J/Q < J_c/Q$, and we show that long but finite antiferromagnetic correlations 
are still present.

\end{abstract}

\newpage

\section{Introduction}

Undoped antiferromagnets, which can be modeled with the spin $\frac{1}{2}$ 
Heisenberg Hamiltonian, are among the quantitatively best understood condensed 
matter systems. To a large extent this is due to an interplay of the very 
efficient loop-cluster algorithm \cite{Eve93,Wie94,Bea96} with the effective
field theory for antiferromagnetic magnons 
\cite{Cha89,Neu89,Fis89,Has90,Has91}. In particular, applying chiral 
perturbation theory --- the systematic low-energy effective field theory for 
Goldstone bosons --- to antiferromagnetic magnons, Hasenfratz and Niedermayer 
\cite{Has93} have derived analytic expressions for the staggered and uniform 
susceptibilities. By comparing these expressions with very accurate Monte Carlo
data obtained with a loop-cluster algorithm, the staggered magnetization, the
spin stiffness, as well as the spinwave velocity of the Heisenberg model have 
been determined very precisely \cite{Wie94,Bea96}. In particular, the resulting
values of these low-energy parameters are in quantitative agreement with 
experimental results on undoped antiferromagnets \cite{Gre94}.

High-temperature superconductors result from doping their antiferromagnetic
precursor insulators. With increased doping, antiferromagnetism is destroyed 
before high-temperature superconductivity emerges. Understanding the doped 
systems from first principles is very difficult because numerical simulations 
of microscopic systems such as the Hubbard or $t$-$J$ model suffer from a 
severe fermion sign problem at a non-zero density of charge carriers. In the 
cuprates, antiferromagnetism and high-temperature superconductivity are 
separated by a pseudo-gap regime. It has been conjectured that this regime is 
connected to a quantum critical point with unusual properties. In particular, 
the N\'eel order of the antiferromagnet may give way to a spin liquid phase 
without long-range magnetic order before the phase coherence of the Cooper 
pairs of high-temperature superconductivity sets in at somewhat larger doping. 

According to the Ginzburg-Landau-Wilson paradigm, a direct phase transition 
separating one type of order from another should generically be of first order.
This paradigm has recently been challenged by the idea of deconfined quantum 
criticality \cite{Sen04}. A deconfined quantum critical point is a second order
phase transition directly separating two competing ordered phases as, for
example, an antiferromagnet or a superfluid from a valence bond solid (VBS).
There are two types for VBS order: columnar and plaquette order, which are 
illustrated in figure 1.
\begin{figure}
\vspace{-2cm}
\begin{center}
\epsfig{file=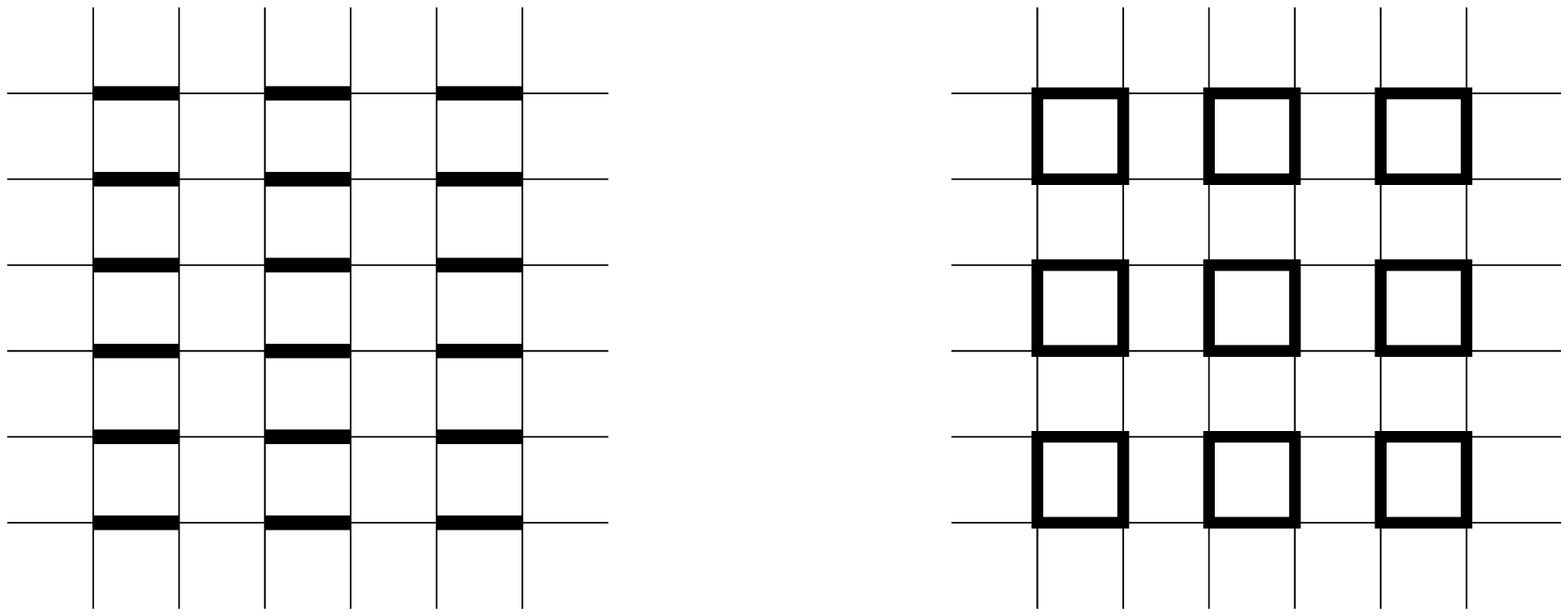,angle=90,width=15cm}
\end{center}
\vspace{-4cm}
\caption{\it Columnar (a) and plaquette (b) type of VBS order. The solid bonds
indicate groups of spins that preferentially form singlets.}
\end{figure}
At a deconfined quantum critical point, spinons --- i.e.\ neutral spin
$\frac{1}{2}$ excitations which are 
confined in the two ordered phases --- are liberated and exist as deconfined 
physical degrees of freedom. It was conjectured that the continuum field theory
that describes a deconfined quantum critical point separating an 
antiferromagnet from a VBS state should be a $(2+1)$-dimensional $\CP(1)$ model
with a dynamical non-compact $U(1)$ gauge field. This theory is expected to be
in the same universality class as an $O(3)$ non-linear $\sigma$-model in which
the creation or annihilation of baby-Skyrmions is forbidden \cite{Mot03}. The
resulting conserved number of baby-Skyrmions gives rise to an additional $U(1)$
symmetry. In \cite{Kau07} it has been argued that --- upon doping --- a
deconfined quantum critical point separating antiferromagnetism from VBS order
may extend to a spin liquid phase, thus providing a possible explanation for
the pseudo-gap regime in under-doped cuprates. 

Establishing the existence of deconfined quantum criticality in an actual
physical system is a non-trivial issue. For example, numerical simulations of
microscopic models with a transition separating superfluidity from VBS order 
found a weak first order transition \cite{Mel04,Kuk04}. Using their flowgram 
method, in a detailed study of another superfluid-VBS transition Kuklov, 
Prokof'ev, Svistunov, and Troyer have again established a weak first order 
transition instead of a quantum critical point \cite{Kuk06}, thus confirming 
the Ginzburg-Landau-Wilson paradigm also in that case.

While unbiased numerical simulations of sufficiently strongly doped 
antiferromagnets are prevented by a severe fermion sign problem, Sandvik has 
pointed out that there is no sign problem in the spin $\frac{1}{2}$ Heisenberg
model with a particular four-spin interaction similar to ring-exchange
\cite{San06}. For this model, he presented numerical evidence for a deconfined
quantum critical point separating antiferromagnetism from VBS order. The
quantum Monte Carlo study of \cite{San06} was performed using a projector Monte
Carlo method in the valence bond basis \cite{San05,San07} and was limited to 
zero 
temperature and to moderate volumes. Recently, Melko and Kaul have simulated 
the same system on larger lattices at finite temperature using a stochastic 
series expansion method \cite{Mel07}. Both studies \cite{San06,Mel07} conclude 
that the transition belongs to a new universality class that is inconsistent 
with the Ginzburg-Landau-Wilson paradigm. As we will discuss, this conclusion 
rests on the use of sub-leading corrections to scaling, which can, however, not
be determined unambiguously from the data. In this paper, we apply a rather
efficient loop-cluster algorithm to the same system. This has allowed us to
also reach large volumes. In order to decide if the transition is second or
weakly first order, we have implemented the flowgram method of \cite{Kuk06}.
Our data provide evidence that the transition is weakly first order, i.e.\ the 
Ginzburg-Landau-Wilson paradigm is again confirmed. This means that an $SU(2)$
or $U(1)$ invariant system, for which the phenomenon of deconfined quantum 
criticality can be firmly established, has yet to be found. 
\footnote{Renormalization group arguments have been used to demonstrate the 
existence of deconfined quantum criticality in an $SU(N)$-invariant system at
very large $N$.} Finding
such a system is non-trivial, in particular, since it should be accessible to
accurate first principles numerical simulations. In this context, it is 
interesting to consult \cite{Har07,Gro07}.

Besides studying the phase transition, we also investigate in detail how 
antiferromagnetism is weakened. In particular, we extend the results of 
\cite{Wie94,Bea96} by determining the staggered magnetization, the spin 
stiffness, as well as the spinwave velocity of the Heisenberg antiferromagnet 
as functions of the strength of the four-spin interaction. In addition, we 
investigate some properties of the VBS phase.

The rest of the paper is organized as follows. The Heisenberg model with 
four-spin interactions as well as some relevant observables are introduced in
section 2. In section 3 the weakening of antiferromagnetism is studied by 
comparing Monte Carlo data with analytic predictions from the systematic 
low-energy effective theory for magnons. In section 4 the phase transition and,
in particular, the question of its order is investigated. Some properties of 
the VBS phase are studied in section 5. Finally, section 6 contains our
conclusions.

\section{Spin Model and Observables}

In this section we introduce the microscopic Heisenberg Hamiltonian with 
four-spin interaction, as well as some relevant observables.

\subsection{Heisenberg Model with Four-Spin Interaction}

Let us consider the spin $\frac{1}{2}$ Heisenberg model on a 2-dimensional 
periodic square lattice of side length $L$ with an additional four-spin 
interaction defined by the Hamiltonian
\begin{eqnarray}
H = J \sum_{x,i} \vec S_x \cdot \vec S_{x+\hat i} - Q \sum_x\!\!\!\!\!\!\!\!\!
&&\left[(\vec S_x \cdot \vec S_{x + \hat 1} - \frac{1}{4})
(\vec S_{x + \hat 2} \cdot \vec S_{x + \hat 1 + \hat 2} - \frac{1}{4}) \right.
\nonumber \\
&&\left.+ (\vec S_x \cdot \vec S_{x + \hat 2} - \frac{1}{4})
(\vec S_{x + \hat 1} \cdot \vec S_{x + \hat 1 + \hat 2} - \frac{1}{4})\right].
\end{eqnarray}
Here $\vec S_x = \frac{1}{2} \vec \sigma_x$ is a spin $\frac{1}{2}$ operator 
located at the lattice site $x$ and $\hat i$ is a vector of length $a$ (where 
$a$ is the lattice spacing) pointing in the $i$-direction. The standard 
exchange coupling $J > 0$ favors anti-parallel spins. The four-spin coupling 
$Q > 0$ favors the simultaneous formation of singlet pairs on opposite sides of
an elementary plaquette. Sandvik has pointed out that quantum Monte Carlo 
simulations of this four-spin interaction do not suffer from the sign problem 
\cite{San06}. Indeed, the fact that it can be treated reliably in numerical 
simulations is the main reason to consider this particular form of the 
coupling.

\subsection{Observables}

Obviously, the above Hamiltonian commutes with the uniform magnetization
\begin{equation}
\vec M = \sum_x \vec S_x. 
\end{equation}
The order parameter for antiferromagnetism is the staggered magnetization
\begin{equation}
\vec M_s = \sum_x (-1)^{(x_1+x_2)/a} \vec S_x.
\end{equation}
A physical quantity of central interest is the staggered susceptibility
\begin{equation}
\chi_s = \frac{1}{L^2} 
\int_0^\beta dt \ \langle M^3_s(0) M^3_s(t) \rangle =
\frac{1}{L^2} \int_0^\beta dt \ \frac{1}{Z} 
\mbox{Tr}[M^3_s(0) M^3_s(t) \exp(- \beta H)],
\end{equation}
the integrated correlation function of the 3-component of the staggered 
magnetization operator. Here $\beta$ is the inverse temperature and
\begin{equation}
Z = \mbox{Tr}\exp(- \beta H)
\end{equation}
is the partition function. Another relevant quantity is the uniform
susceptibility
\begin{equation}
\chi_u = \frac{1}{L^2} \int_0^\beta dt \ 
\langle M^3(0) M^3(t) \rangle =
\frac{1}{L^2} \int_0^\beta dt \ \frac{1}{Z} \mbox{Tr}[M^3(0) M^3(t)
\exp(- \beta H)],
\end{equation}
the integrated correlation function of the uniform magnetization. Both $\chi_s$
and $\chi_u$ can be measured very efficiently with the loop-cluster algorithm
using improved estimators \cite{Wie94}. In particular, in the multi-cluster
version of the algorithm the staggered susceptibility
\begin{equation}
\chi_s = \frac{1}{4 \beta L^2} \left\langle \sum_{\cal C} |{\cal C}|^2 
\right\rangle
\end{equation}
is given in terms of the cluster sizes $|{\cal C}|$ (which have the dimension
of time). Similarly, the uniform susceptibility
\begin{equation}
\chi_u = \frac{\beta}{4 L^2} \left\langle W_t^2 \right\rangle =
\frac{\beta}{4 L^2} \left\langle \sum_{\cal C} W_t({\cal C})^2 
\right\rangle
\end{equation}
is given in terms of the temporal winding number
$W_t = \sum_{\cal C} W_t({\cal C})$, which is the sum of winding numbers
$W_t({\cal C})$ of the loop-clusters ${\cal C}$ around the Euclidean time 
direction. In complete analogy, the spatial winding numbers 
$W_i = \sum_{\cal C} W_i({\cal C})$ define two spatial susceptibilities 
\begin{equation}
\chi_i = \frac{1}{4 \beta} \left\langle W_i^2 \right\rangle =
\frac{1}{4 \beta} \left\langle \sum_{\cal C} W_i({\cal C})^2 \right\rangle.
\end{equation}
These susceptibilities measure the response of the system to a twist in the 
spatial boundary conditions.

A natural order parameter that signals a VBS state is
\begin{equation}
D_i = \sum_x (-1)^{x_i/a} \vec S_x \cdot \vec S_{x+\hat i}.
\end{equation}
In a VBS state with columnar order either $D_1$ or $D_2$ has a non-vanishing
vacuum expectation value. In a VBS state with plaquette order, on the other 
hand, one of the linear combinations $D_1 \pm D_2$ has a non-zero expectation 
value. In numerical simulations, it is easier to investigate an alternative 
pair of order parameters which just count the number of spin flips in the 
configurations contributing to the path integral. We define the order parameter
$\widetilde D_i$ as the difference between the number of spin flips on 
nearest-neighbor bonds in the $i$-direction with an even and an odd value of 
$x_i/a$. It should be noted that such flips can be due to both the standard 
two-spin coupling of strength $J$ and the four-spin coupling of strength $Q$. 
The corresponding probability distribution $p(\widetilde D_1,\widetilde D_2)$
is useful for investigating the nature of the VBS state.

\section{Weakening of Antiferromagnetism}

In this section we investigate the weakening of antiferromagnetism. First, we
briefly review some results of the systematic low-energy magnon effective field
theory. Then Monte Carlo data obtained with a loop-cluster algorithm are used 
to determine the values of the low-energy parameters of the effective theory.

\subsection{Low-Energy Effective Theory for Magnons}

The low-energy physics of antiferromagnets is determined by the $SU(2)_s$ spin 
symmetry which is spontaneously broken down to $U(1)_s$.
As a result, there are two massless Goldstone bosons --- the antiferromagnetic 
spinwaves  or magnons. Chakravarty, Halperin, and Nelson \cite{Cha89} were 
first to describe the low-energy magnon physics by an effective field theory 
--- the $(2+1)$-d $O(3)$-invariant non-linear $\sigma$-model. In analogy to 
chiral perturbation theory for the pseudo-Goldstone pions in QCD, a systematic 
low-energy effective field theory for magnons was developed in 
\cite{Neu89,Fis89,Has90,Has91}. The staggered magnetization of an 
antiferromagnet is described by a unit-vector field
\begin{equation}
\vec e(x) = (e_1(x),e_2(x),e_3(x)), \qquad \vec e(x)^2 = 1,
\end{equation}
in the coset space $SU(2)_s/U(1)_s = S^2$. Here $x = (x_1,x_2,t)$ denotes
a point in space-time. To leading order, the Euclidean magnon effective action 
takes the form
\begin{equation}
S[\vec e] = \int d^2x \ dt \ \frac{\rho_s}{2} 
\left(\p_i \vec e \cdot \p_i \vec e +
\frac{1}{c^2} \p_t \vec e \cdot \p_t \vec e\right).
\end{equation}
The index $i \in \{1,2\}$ labels the two spatial directions, while the index 
$t$ refers to the Euclidean time-direction. The parameter $\rho_s$ is the spin 
stiffness and $c$ is the spinwave velocity. At low energies the antiferromagnet
has a relativistic spectrum. Hence, by introducing $x_0 = c t$ the action takes
the manifestly Lorentz-invariant form
\begin{equation}
S[\vec e] = \int d^2x \ dx_0 \ \frac{\rho_s}{2 c} 
\p_\mu \vec e \cdot \p_\mu \vec e.
\end{equation}
The ratio $\xi = c/(2 \pi \rho_s)$ defines a characteristic length scale which 
diverges when antiferromagnetism disappears at a second order phase transition.

Hasenfratz and Niedermayer have performed very detailed calculations of a 
variety of physical quantities including the next to next to leading 2-loop 
order of the systematic expansion \cite{Has93}. For our study their results for
finite temperature and finite volume effects of the staggered and uniform
susceptibilities are most relevant. Depending on the size $L$ of the quadratic
periodic spatial volume and the inverse temperature $\beta$, one distinguishes
cubical space-time volumes with $L \approx \beta c$ from cylindrical ones with
$\beta c \gg L$. The aspect ratio of the space-time box is characterized by
\begin{equation}
l = \left({\frac{\beta c}{L}}\right)^{1/3}.
\end{equation}
In the cubical regime the volume- and temperature-dependence of the staggered 
magnetization is given by
\begin{equation}
\label{chiscube}
\chi_s = \frac{{\cal M}_s^2 L^2 \beta}{3} 
\left\{1 + 2 \frac{c}{\rho_s L l} \beta_1(l) +
\left(\frac{c}{\rho_s L l}\right)^2 \left[\beta_1(l)^2 + 3 \beta_2(l)\right] + 
...\right\},
\end{equation}
where ${\cal M}_s$ is the staggered magnetization density. The uniform 
susceptibility takes the form
\begin{equation}
\label{chiucube}
\chi_u = \frac{2 \rho_s}{3 c^2} 
\left\{1 + \frac{1}{3} \frac{c}{\rho_s L l} \widetilde\beta_1(l) +
\frac{1}{3} \left(\frac{c}{\rho_s L l}\right)^2 
\left[\widetilde\beta_2(l) - \frac{1}{3} \widetilde\beta_1(l)^2 - 6 \psi(l)
\right] + ...\right\}.
\end{equation}
The functions $\beta_i(l)$, $\widetilde\beta_i(l)$, and $\psi(l)$ are shape 
coefficients of the space-time box defined in \cite{Has93}.

\subsection{Determination of the Low-Energy Parameters}

We have performed numerical simulations of the Heisenberg model with four-spin
interaction for a variety of lattice sizes $L/a$ ranging from $24$ to $112$ 
at inverse temperatures between $\beta J = 10$ and 20. Remarkably, just like
the ordinary two-spin coupling, the additional four-spin coupling can also be
treated with an efficient loop-cluster algorithm. The algorithm, presently
implemented only in discrete time, will be described elsewhere. All simulations
described in this section have been performed at three different lattice 
spacings in discrete time, which allows a reliable extrapolation to the 
continuum limit. Some numerical data (extrapolated to the time-continuum limit)
are listed in table 1. 
\begin{table}
\begin{center}
\begin{tabular}{|c|c|c|c|c|}
\hline 
$Q/J$&
$\beta J$&
$L/a$&
$\chi_s J a$&
$\langle W_t^2 \rangle$\tabularnewline
\hline
\hline 
$0.1$&
$20$&
$34$&
$743.0(1.8)$&
$8.285(26)$\tabularnewline
\hline 
$0.1$&
$20$&
$36$&
$829(2)$&
$9.312(25)$\tabularnewline
\hline 
$0.5$&
$16$&
$42$&
$625.1(1.6)$&
$7.561(20)$\tabularnewline
\hline 
$0.5$&
$16$&
$44$&
$683(2)$&
$8.310(20)$\tabularnewline
\hline 
$1$&
$12$&
$42$&
$333.7(1.3)$&
$5.361(20)$\tabularnewline
\hline 
$1$&
$12$&
$46$&
$396.3(1.5)$&
$6.434(22)$\tabularnewline
\hline 
$2$&
$12$&
$66$&
$470.8(1.6)$&
$5.598(20)$\tabularnewline
\hline 
$2$&
$12$&
$68$&
$497.7(1.6)$&
$5.960(22)$\tabularnewline
\hline 
$3$&
$10$&
$78$&
$383.8(1.2)$&
$5.310(21)$\tabularnewline
\hline 
$3$&
$10$&
$82$&
$420.8(1.2)$&
$5.914(22)$\tabularnewline
\hline 
$4$&
$10$&
$94$&
$415.9(1.5)$&
$5.011(23)$\tabularnewline
\hline 
$4$&
$10$&
$96$&
$431.4(1.5)$&
$5.229(26)$\tabularnewline
\hline
\end{tabular}
\end{center}
\caption{\it Some numerical data for the staggered susceptibility and the
temporal winding number squared $\langle W_t^2 \rangle$ obtained with the
loop-cluster algorithm.}
\end{table}
For fixed $J$ and $Q$ all data for $\chi_s$ and $\chi_u$ have been fitted 
simultaneously to eqs.(\ref{chiscube}) and (\ref{chiucube}) by using the
low-energy constants ${\cal M}_s$, $\rho_s$, and $c$ as fit parameters. The 
fits are very good with $\chi^2/\mbox{d.o.f.}$ ranging from 0.5 to 2.0. Typical
fits are shown in figures 2a and 2b. 
\begin{figure}
\begin{center}
\vspace{-2cm}
\epsfig{file=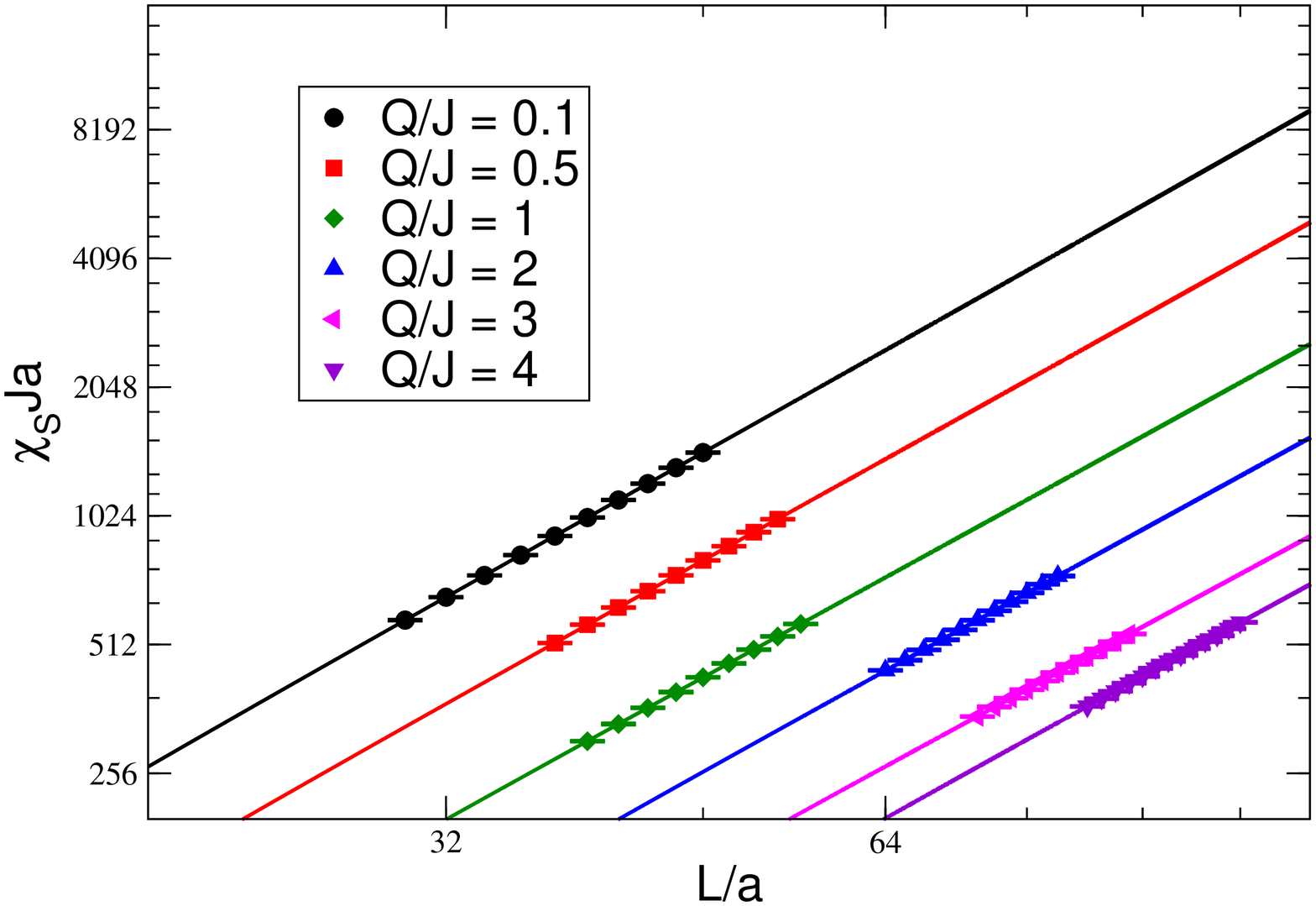,angle=-90,width=12cm} \vskip-0.5cm
\epsfig{file=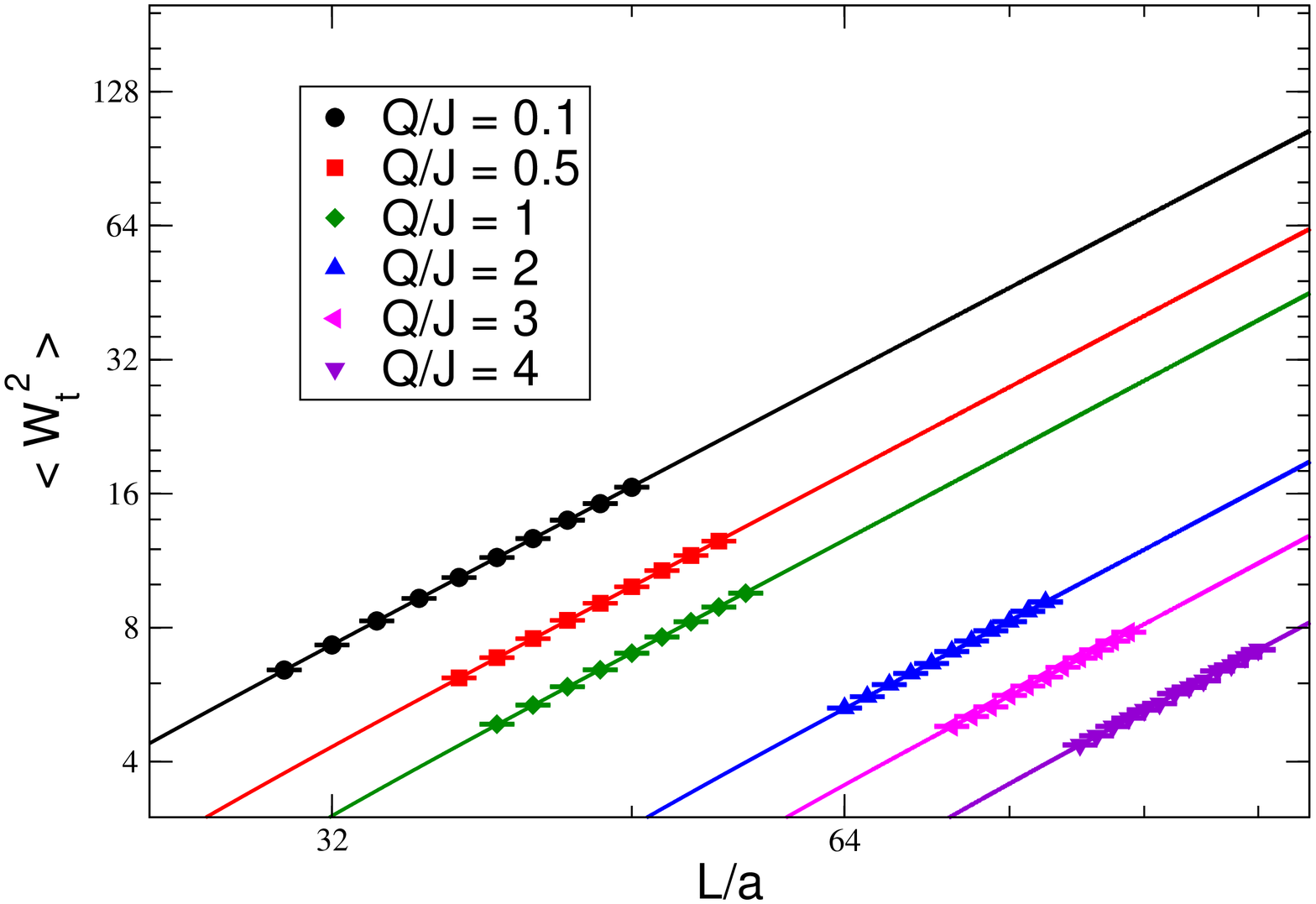,angle=-90,width=12cm} \vskip-0.5cm
\end{center}
\caption{\it Fit of the finite-size and finite-temperature effects of the
staggered susceptibility $\chi_s$ (a) and the temporal winding number squared
$\langle W_t^2 \rangle$ (b) to results of the effective theory in the cubical 
regime for various values of $Q/J$.}
\end{figure}
The corresponding results are summarized in table 2 and illustrated in figures
3 and 4. One observes a substantial weakening of antiferromagnetism. In
particular, as one goes from $Q = 0$ to $Q = 4 J$, the staggered magnetization
${\cal M}_s$ decreases by a factor of about 3, while the correlation length
$\xi = c/(2 \pi \rho_s)$ increases by a factor of about 5. Interestingly, in 
units of $J$, the spin stiffness $\rho_s$ is more or less constant. The 
increase of $\xi$ with $Q$ is thus due to an increase of the spinwave velocity
$c$ (in units of $Ja^2$). When antiferromagnetism disappears at a second order
phase transition, the correlation length $\xi$ diverges. This is possible, only
if $\rho_s$ goes to zero at the transition. Since the system interacts locally,
any excitation travels with a finite speed, and hence $c$ cannot go to
infinity. In the next section we will present numerical evidence for a first
order phase transition. In that case, $\rho_s$ remains finite at the transition.
\begin{table}
\begin{center}
\begin{tabular}{|c|c|c|c|c|}
\hline
$Q/J$ & ${\cal M}_s a^2$ & $\rho_s/J$ & $c/(Ja^2)$ & $\xi/a$ \\
\hline
\hline
0   & 0.3074(4) & 0.186(4) & 1.68(1) & 1.44(3) \\
\hline
0.1 & 0.2909(6) & 0.183(6) & 1.88(3) & 1.64(3) \\
\hline
0.5 & 0.2383(7) & 0.182(6) & 2.73(4) & 2.39(4) \\
\hline
1   & 0.1965(7) & 0.194(7) & 3.90(6) & 3.19(5) \\
\hline
2   & 0.149(1)  & 0.194(9) & 5.98(14) & 4.91(12) \\
\hline
3   & 0.122(1)  & 0.192(8) & 7.97(16) & 6.60(14) \\
\hline
4   & 0.106(1)  & 0.218(13) & 10.50(31) & 7.67(26) \\
\hline
\end{tabular}
\end{center}
\caption{\it Results for the low-energy parameters ${\cal M}_s$, $\rho_s$, and
$c$ as well as the length scale $\xi = c/(2 \pi \rho_s)$ obtained from fitting 
$\chi_s$ and $\chi_u$ to the analytic expressions of eqs.(\ref{chiscube}) and
(\ref{chiucube}) from the magnon effective theory.}
\end{table}
\begin{figure}
\begin{center}
\epsfig{file=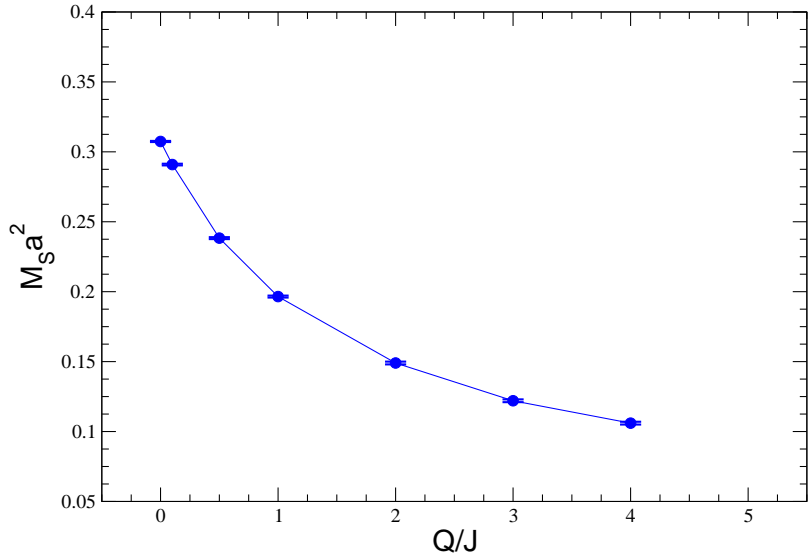,angle=-90,width=12cm}
\end{center}
\caption{\it The staggered magnetization ${\cal M}_s$ as a function of $Q/J$,
obtained from the fits to the magnon effective theory results for $\chi_s$ 
and $\chi_u$.}
\end{figure}
\begin{figure}
\begin{center}
\epsfig{file=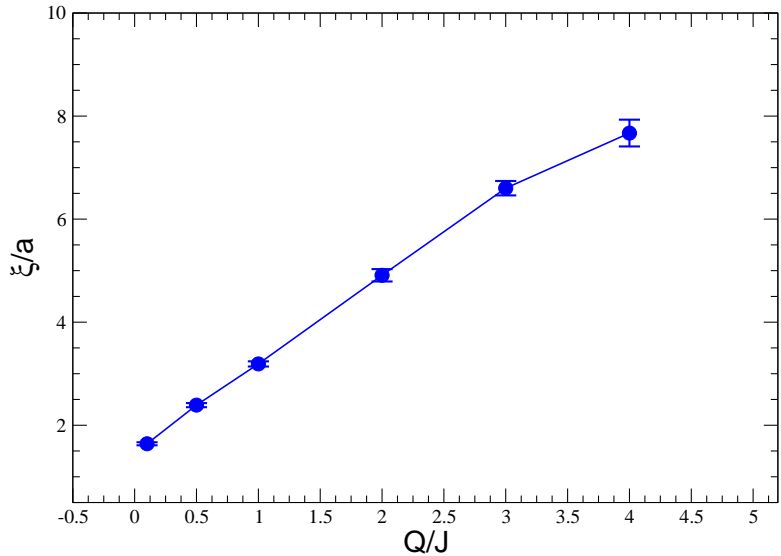,angle=-90,width=12cm}
\end{center}
\caption{\it The length scale $\xi = c/(2 \pi \rho_s)$ as a function of $Q/J$,
obtained from the fits to the magnon effective theory results for $\chi_s$ 
and $\chi_u$.}
\end{figure}

\section{Phase Transition between Antiferromagnetism and VBS Order}

In this section we study the phase transition at which antiferromagnetism
turns into VBS order. In particular, the order of the transition is 
investigated using both finite-size scaling and the flowgram method of 
\cite{Kuk06}.

\subsection{Finite-Size Effects of $\langle W_i^2 \rangle$ Near the Transition}

As we have seen in the previous section, antiferromagnetism is substantially
weakened as the four-spin coupling $Q$ increases. This manifests itself in the
reduction of the staggered magnetization ${\cal M}_s$ as well as in the 
increase of the characteristic length scale $\xi = c/(2 \pi \rho_s)$. The 
higher order terms in the systematic expansion are suppressed as long as 
$L \gg \xi$. In practice, this limits us to $\xi \approx 10 a$ which 
corresponds to $Q/J \approx 5$. As one approaches a second order phase 
transition, $\xi$ diverges and the systematic effective theory is no longer 
applicable. Instead, in the vicinity of the phase transition, it is useful to 
employ finite-size scaling. 

In order to locate the transition it is natural to investigate the 
$J/Q$-dependence of the spatial winding number squared $\langle W_i^2 \rangle$.
In particular, in case of a second order phase transition, for sufficiently
large volumes the various finite volume curves should all intersect at the
critical coupling. Recently, such an analysis has been reported by Melko and
Kaul \cite{Mel07}. We have verified explicitly that our Monte Carlo data are
consistent with those of that study. In figure 5a we show a fit to those data 
for moderate volumes $L/a = 32,40$, and 48 near the transition using the 
finite-size scaling ansatz
\begin{equation}
\label{fit}
\langle W_i^2 \rangle = f\left(\frac{J - J_c}{J_c} L^{1/\nu}\right) =
A + B \frac{J - J_c}{J_c} L^{1/\nu} + 
{\cal O}\left(\left(\frac{J - J_c}{J_c}\right)^2\right).
\end{equation}
The fit is good, with $\chi^2/\mbox{d.o.f.} \approx 2$, suggesting that the 
transition might actually be second order. In particular, the three finite 
volume curves intersect in one point, $J_c/Q = 0.0375(5)$, and do not require 
an additive sub-leading correction $C L^{-\omega}$ to eq.(\ref{fit}).
\begin{figure}
\begin{center}
\vspace{-2cm}
\epsfig{file=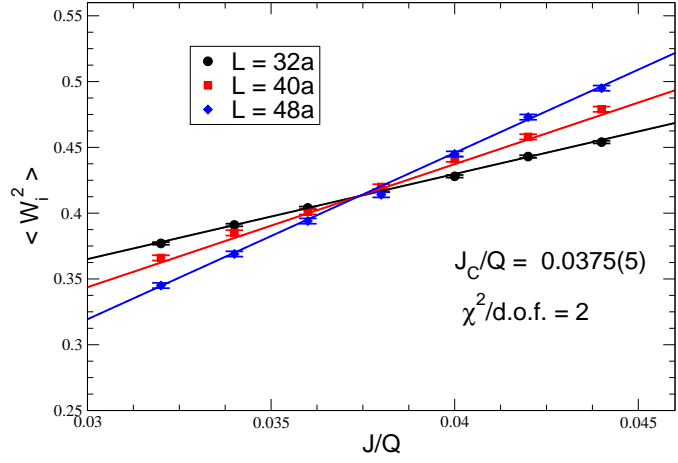,angle=-90,width=10cm} \vskip-0.5cm
\epsfig{file=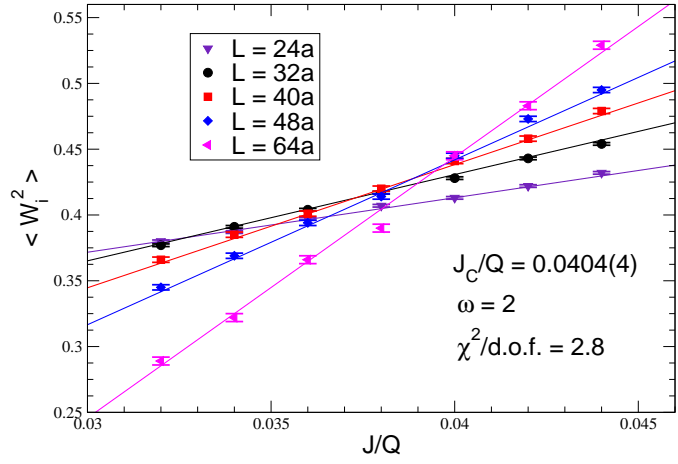,angle=-90,width=10cm} \vskip-0.5cm
\epsfig{file=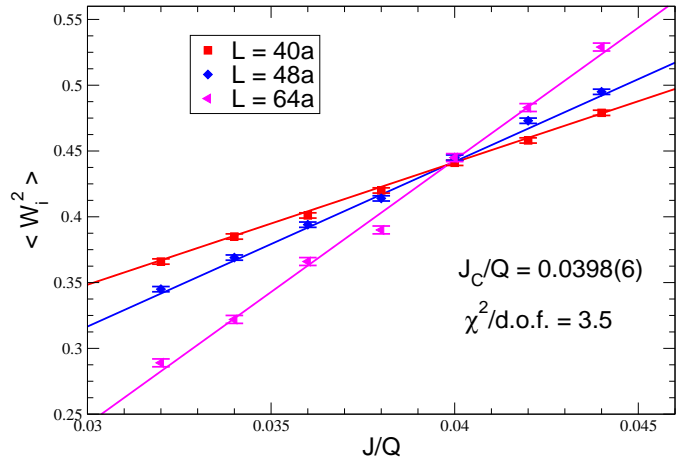,angle=-90,width=10cm} \vskip-0.5cm
\caption{\it Three different fits of the spatial winding number squared
$\langle W_i^2 \rangle$ as a function of the coupling $J/Q$ in
the transition region.}
\end{center}
\end{figure}
This is consistent 
with Sandvik's earlier result obtained on smaller volumes $L/a = 16,...,32$ 
which did require the inclusion of the sub-leading term. His fit led to 
$\omega \approx 2$ which implies that the corrections are suppressed for large 
volumes. Remarkably, when Melko and Kaul's $L/a = 64$ data are included in the
fit of eq.(\ref{fit}), its quality degrades to $\chi^2/\mbox{d.o.f.} \approx 8$.
In fact, the $L/a = 64$ curve does not pass through the intersection point of
the smaller volume curves. Melko and Kaul attribute this behavior to 
sub-leading terms and finally come to the conclusion that there is a deconfined
quantum critical point somewhere in the interval $0.038 < J/Q < 0.040$. Indeed,
a fit including an additional sub-leading term $C L^{-\omega}$ is possible. 
However, the exponent $\omega$ is not well determined by the data. In order to
obtain a stable fit, we have fixed $\omega$ to different values ranging from 
0.01 to 2.5, which all give more or less the same $\chi^2/\mbox{d.o.f.} \approx 
2.5$, but lead to different values of the critical coupling $J_c$. For 
example, fixing $\omega = 2$, as suggested by \cite{San06}, one obtains 
$J_c/Q = 0.0404(4)$ and $\nu = 0.62(2)$. This fit is illustrated in figure 5b. 
On the other hand, when one fixes $\omega = 0.01$, one obtains 
$J_c/Q = 0.0438(7)$ and $\nu = 0.62(2)$. Finally, when one excludes all but the 
largest volumes $L/a = 40,48$, and 64, a four-parameter fit becomes possible 
again. This fit, shown in figure 5c, is less good with 
$\chi^2/\mbox{d.o.f.} \approx 3.5$ and it yields $J_c/Q = 0.0398(6)$, which is 
inconsistent with the critical coupling obtained from the moderate volume data.
Even larger volumes would be needed in order to decide if the curves will 
continue to intersect in the same point. 

To summarize, the moderate volume data (with $L/a = 32,40$, and 48) are well 
described by the four-parameter fit of eq.(\ref{fit}), while all data including
those for $L/a = 64$ are not. These data can be described by a six-parameter
fit including sub-leading corrections, but the data do not unambiguously
determine the fit parameters. Since no sub-leading term is required to fit the
moderate volume data, it seems strange that such corrections become necessary
once larger volumes are included in the fit. We take this unusual behavior as a
first indication that the transition may actually be weakly first order.

Another observation that may cast some doubt on the second order nature of the 
transition is a non-monotonic behavior of $\langle W_i^2 \rangle$ near the
transition, which is displayed in figure 6. 
\begin{figure}
\begin{center}
\epsfig{file=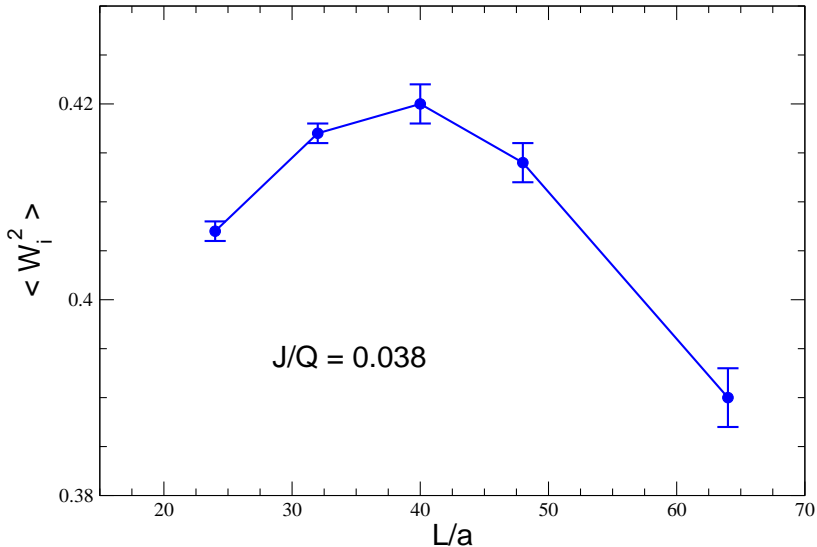,angle=-90,width=12cm}
\end{center}
\caption{\it Non-monotonic volume-dependence of $\langle W_i^2 \rangle$ at 
$J/Q = 0.038$ near the critical coupling that may indicate a weak first order 
phase transition.}
\end{figure}
Such behavior is typical for a first order phase transition. For example, in
the VBS phase, at a point close to a first order transition, domains of 
antiferromagnetic phase can still exist. Thus, for small volumes, the
antiferromagnetic domains may lead to a linear increase of 
$\langle W_i^2 \rangle$ with $L$. For larger volumes, the VBS phase will begin
to dominate and $\langle W_i^2 \rangle$ will then decrease. This competition can
lead to non-monotonic behavior. For these reasons, we think that the
data of \cite{San06} and \cite{Mel07} do not provide sufficiently convincing 
evidence that deconfined quantum criticality has actually been observed. Due to
limited numerical resources, we have not been able to extend the analysis to 
substantially larger volumes. However, using a supercomputer this would 
definitely be possible and, in fact, highly desirable. In order to shed more 
light on the subtle issue of quantum criticality versus a weak first order 
transition, we now turn to an alternative method of analysis. 

\subsection{Application of the Flowgram Method}

Kuklov, Prokof'ev, Svistunov, and Troyer have developed a flowgram method
which is useful for distinguishing weakly first order from second order phase 
transitions \cite{Kuk06}. For our system, the flowgram method can be 
implemented as follows. We work on lattices of increasing size $L$ at the 
inverse temperature given by $\beta Q = L/a$. First, each individual spin 
configuration in the path integral
is associated with either the antiferromagnetic or the VBS phase according to
the following criterion. If all three winding numbers $W_1$, $W_2$, and $W_t$
are equal to zero, the configuration is associated with the VBS phase. On the 
other hand, if at least one of the three winding numbers is non-zero, the
configuration is associated with the antiferromagnetic phase. This criterion is
natural, because in the infinite volume limit there is no winding in the VBS
phase, while there is always some winding in the antiferromagnetic phase. One
then defines a volume-dependent pseudo-critical coupling $J_c(L)$ at which both
competing phases have equal weight, i.e.\ the number of associated 
configurations is the same for both phases. It is important to note that, in 
the infinite volume limit, the pseudo-critical coupling $J_c(L)$ approaches the
true location of the phase transition both for a first and for a second order 
phase transition. The large volume limit is now approached by simulating 
systems at the pseudo-critical coupling $J_c(L)$ for increasing values of $L$. 
Defining the sum of the spatial and temporal winding numbers squared as
$W^2 = W_1^2 + W_2^2 + W_t^2$, the quantity $\langle W^2 \rangle(J_c(L))$ is
then evaluated at the pseudo-critical coupling $J_c(L)$. If the phase 
transition is second order, $\langle W^2 \rangle(J_c(L))$ will approach a 
constant for large $L$ since $\rho_s$ then vanishes (i.e.\ 
$\xi = c/(2 \pi \rho_s)$ diverges) at the transition. On the other hand, if the
transition is first order, with 50 percent probability the system still shows 
the characteristics of the antiferromagnet. Thus, for $L \gg \xi$, 
$\langle W^2 \rangle(J_c(L))$ grows linearly with $L$. As we will see below, 
for $48a \leq L \leq 80a$ we indeed observe this behavior.

We have implemented the Ferrenberg-Swendsen re-weighting method \cite{Fer88} in
order to accurately locate the pseudo-critical coupling. Unlike in the rest of
this paper, the simulations in this subsection have only been performed at two
(instead of three) lattice spacings in discrete time. Both lattice spacings are
close to the continuum limit and give consistent results. Instead of 
extrapolating to the continuum limit (which is less reliable with two than with
three lattice spacings), in this subsection we quote our results at the smaller
lattice spacing $\varepsilon Q = 0.05$. A calculation closer to the continuum
limit or, even better, using a continuous-time algorithm would still be useful.

In order to investigate whether the phase transition is second or weakly first
order, the values of $\langle W^2 \rangle$ at the pseudo-critical coupling 
$J_c(L)$ are illustrated in figure 7.
\begin{figure}
\begin{center}
\epsfig{file=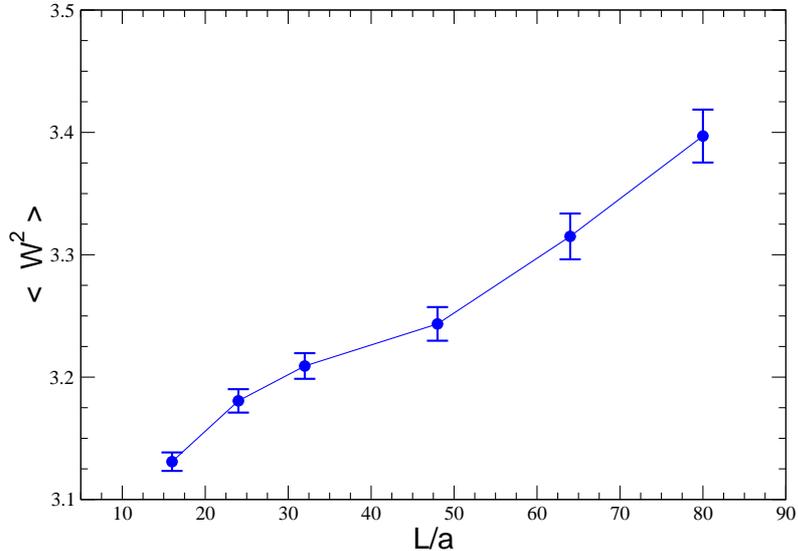,angle=270,width=12cm}
\end{center}
\caption{\it The sum of spatial and temporal winding numbers squared 
$\langle W^2 \rangle(J_c(L))$ evaluated at the pseudo-critical coupling 
$J_c(L)$ for increasing lattice size $L$.}
\end{figure}
For moderate volumes up to $L = 48a$ the curve seems to level off, which would
be characteristic of a second order phase transition. Indeed, as we have seen
before, the moderate volume data for the spatial susceptibility are consistent
with the finite-size scaling behavior of a second order phase transition.
However, for larger volumes the curve rises linearly, thus indicating a weak 
first order phase transition. Of course, one cannot completely exclude that the
curve may eventually level off at even larger volumes. We find this unlikely 
and conclude that our results cast serious doubt on the picture of deconfined 
quantum criticality painted in the earlier studies. 

Given the evidence for a weak first order transition, we like to determine the 
value of the critical coupling $J_c$ in the infinite volume. The values of the
pseudo-critical coupling $J_c(L)$ in a finite volume are summarized in table 3.
\begin{table}
\begin{center}
\begin{tabular}{|c|c|c|c|c|c|c|}
\hline
$L/a$ & 24 & 32 & 48 & 64 & 80 & 96 \\
\hline
\hline
$J_c(L)/Q$ & 0.0311(4) & 0.0316(3) & 0.0337(4) & 0.0364(3) & 0.0384(3) & --- \\
\hline
$J_c'(L)/Q$ & 0.115(2) & 0.0871(4) & 0.0632(4) & 0.0544(5) & 0.0477(4) &
0.0445(4) \\
\hline
\end{tabular}
\end{center}
\caption{\it Values of the volume-dependent pseudo-critical couplings $J_c(L)$
and  $J_c'(L)$ obtained with the Ferrenberg-Swendsen re-weighting method.}
\end{table}
Given the data for $J_c(L)$ alone, it is non-trivial to extract the infinite 
volume critical coupling $J_c = J_c(L \rightarrow \infty)$. For this reason, we
have defined another pseudo-critical coupling $J_c'(L)$, which also 
extrapolates to the correct limit, i.e.\ $J_c'(L \rightarrow \infty) = J_c$. In
this case, we work at the inverse temperature given by $\beta Q = L/4a$. 
Irrespective of the spatial winding numbers, if the temporal winding number 
$W_t$ is equal to zero, the configuration is now associated with the VBS phase.
On the other hand, if $W_t$ is non-zero, the configuration is associated with 
the antiferromagnetic phase. As before, we define the volume-dependent 
pseudo-critical coupling $J_c'(L)$ such that both phases have equal weight. The
values of $J_c'(L)$ (again quoted at $\varepsilon Q = 0.05$) are also listed in
table 3. According to the finite-size scaling theory for first order phase
transitions, using $\beta L^2 \propto L^3$, both finite-volume pseudo-critical
couplings should approach their common infinite volume limit $J_c$ as
\begin{equation}
\label{fitJc}
J_c(L) = J_c + C \frac{\log L/a}{L^3}, \ 
J_c'(L) = J_c + C' \frac{\log L/a}{L^3}.
\end{equation}
Interestingly, the two pseudo-critical couplings indeed converge to the same
limit. A fit of $J_c(L)$ and $J_c'(L)$ to eq.(\ref{fitJc}) --- shown in figure
8 --- has $\chi^2/\mbox{d.o.f.} \approx 1.15$ and yields the infinite-volume 
critical coupling $J_c = 0.0396(6)$. Again, only the large-volume data show the 
characteristic behavior of a first order phase transition. 
\begin{figure}
\begin{center}
\epsfig{file=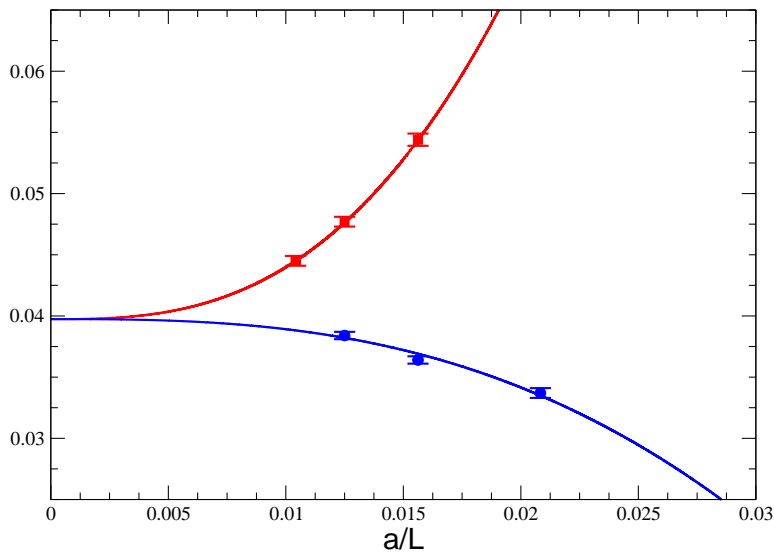,angle=270,width=12cm}
\end{center}
\caption{\it Fit of the pseudo-critical couplings $J_c(L)$ (lower curve) and 
$J_c'(L)$ (upper curve) shown as functions of $a/L$.}
\end{figure}
It should be noted that the definition of $J_c'(L)$ is less natural than the 
one of $J_c(L)$, because it ignores the spatial winding numbers when 
configurations are associated with either of the two phases. In particular, 
$J_c'(L)$ approaches the infinite-volume critical point $J_c$ more slowly than
$J_c(L)$. Consequently, the ultimate large volume physics is more easily 
visible using the pseudo-critical coupling $J_c(L)$. For example, the linear
increase of $\langle W^2 \rangle(J_c(L))$ with $L$, which sets in around 
$L \approx 50a$, is not yet present in $\langle W^2 \rangle(J_c'(L))$, and is 
expected to set in only on larger volumes.

\section{Investigation of the VBS State}

As we have seen, the antiferromagnet is weakened and ultimately destroyed at a 
rather weak first order phase transition. Since the transition is so weak, at 
moderate volumes it is practically indistinguishable from a continuous 
transition. As a result, an approximate $U(1)$ symmetry emerges dynamically as
an enhancement of the discrete 90 degrees rotations of the square lattice. The
other side of the phase transition is characterized by VBS order. However, the 
emergent $U(1)$ symmetry makes it difficult to identify the nature of the VBS 
state as columnar or plaquette.

\subsection{Probability Distribution of the VBS Order Parameter}

In order to investigate the nature of the VBS order it is best to go away 
from the critical point as far as possible (assuming that no other phase 
transitions take place). In the following we thus work at $Q/J = \infty$, which
is obtained by putting $J = 0$. The corresponding probability distribution of 
the standard VBS order parameters $p(D_1,D_2)$ has been determined by Sandvik
for a $32^2$ lattice at zero temperature and it shows perfect $U(1)$ symmetry 
\cite{San06}. The loop-cluster algorithm allows us to repeat this study for 
larger volumes, in this case using the probability distribution of the 
non-standard VBS order parameters $p(\widetilde D_1,\widetilde D_2)$. As one
sees from figure 9, even on a $96^2$ lattice at $\beta Q = 30$ one does not see
any deviation from the $U(1)$ symmetry. Hence, our data do not allow us to
identify the nature of the VBS order.
\begin{figure}
\begin{center}
\epsfig{file=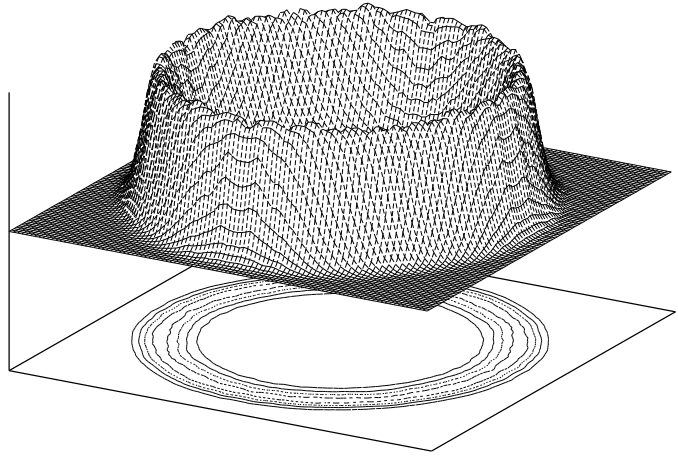,angle=270,width=12cm}
\end{center}
\caption{\it Probability distribution $p(\widetilde D_1,\widetilde D_2)$
obtained on a $96^2$ lattice at $\beta Q = 30$ and $Q/J = \infty$. The observed
$U(1)$ rotation symmetry implies that we cannot identify the nature of the VBS 
phase as either columnar or plaquette.}
\end{figure}

At small $Q/J$, the loop cluster algorithm is extremely efficient with
au\-to-cor\-re\-la\-tions limited to at most a few sweeps. However, at larger 
values of $Q/J$, and especially at $Q/J = \infty$ the algorithm suffers from a 
noticeable auto-cor\-re\-la\-tion problem. This problem arises because the 
cluster algorithm, which is designed to update long-range spin correlations, 
can not efficiently shuffle spin-flip events from even to odd bonds. This 
causes slowing down in the VBS phase. Details concerning the algorithm and its 
performance will be discussed elsewhere.

\subsection{Antiferromagnetic Correlations in the VBS Phase}

In order to confirm that antiferromagnetism indeed disappears for large $Q$,
we again consider $Q/J = \infty$. We have simulated the staggered 
susceptibility as a function of the lattice size $L$. As one sees in figure 10,
at $\beta Q = 50$ the staggered susceptibility $\chi_s$ increases with 
increasing space-time volume until it levels off around $L \approx 50a$. This 
shows that long (but not infinite) range antiferromagnetic correlations survive
even deep in the VBS phase. These data confirm that antiferromagnetism is 
indeed destroyed in the VBS phase. However, again one needs to go to volumes 
larger than $L \approx 50a$ in order to see the ultimate infinite-volume
behavior.
\begin{figure}
\begin{center}
\epsfig{file=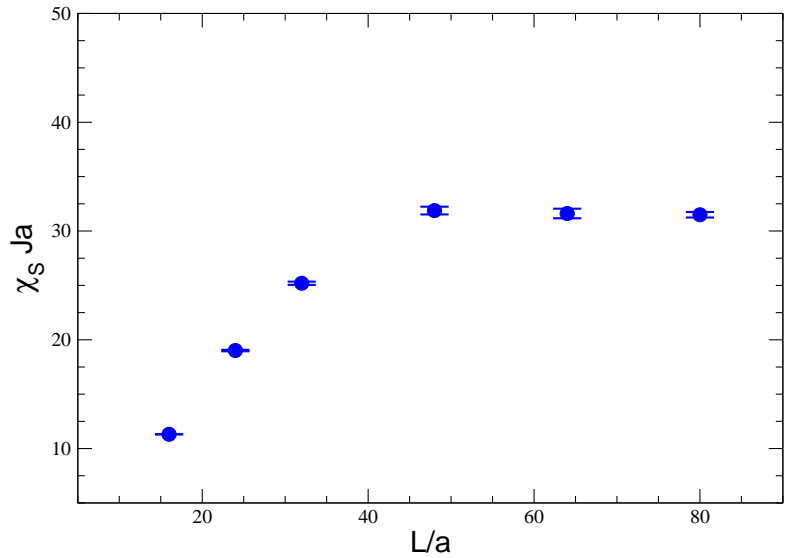,angle=270,width=12cm}
\end{center}
\caption{\it The staggered susceptibility $\chi_s$ in the VBS phase increases 
with increasing space-time volume until it levels off around 
$L/a \approx 50$ for $\beta Q = 50$.}
\end{figure}

\section{Conclusions}

We have employed a rather efficient loop-cluster algorithm to investigate the 
physics of the antiferromagnetic spin $\frac{1}{2}$ Heisenberg model with an
additional four-spin interaction. When the four-spin coupling is sufficiently
strong, antiferromagnetism is destroyed and gives way to a VBS state. While 
Sandvik's pioneering study \cite{San06} was limited to zero temperature and 
moderate volumes, just like the stochastic series expansion method of Melko and
Kaul \cite{Mel07}, the cluster algorithm allows us to work at non-zero
temperatures and large volumes. Using the cluster algorithm and applying the
flowgram method of Kuklov, Prokof'ev, Svistunov, and Troyer \cite{Kuk06}, we 
found numerical evidence for a weak first order phase transition, thus 
supporting the Ginzburg-Landau-Wilson paradigm. Interestingly, the same
conclusion was reached in studies of the transition separating superfluidity 
from VBS order \cite{Mel04,Kuk04,Kuk06}. The first order nature of the phase 
transition in the Heisenberg model with four-spin coupling $Q$ implies that the
idea of deconfined quantum criticality again lacks a physical system for which
it is firmly established. Hence, the proponents of this intriguing idea are 
challenged once more to suggest another microscopic system for which one 
expects this fascinating phenomenon to occur.

It is interesting to ask why the phase transition separating antiferromagnetism
from VBS order is so weakly first order. There must be a reason for the long
correlation length, around $50a$, even if it does not go to infinity. Perhaps,
the ideas behind deconfined quantum criticality may still explain this 
behavior.

\section*{Acknowledgements}

We have benefited from discussions with F.\ Niedermayer and M.\ Troyer. The
work of S.\ C.\ was supported in part by the National Science Foundation under
grant number DMR-0506953. We also acknowledge the support of the
Schweizerischer Nationalfonds.

\end{document}